\documentclass[12pt]{iopart}
\usepackage{lscape}
\usepackage{epsfig}
\usepackage{graphicx}
\usepackage{CJK}
\usepackage{longtable}

\begin{document}
\begin{CJK}{UTF8}{gbsn}

\title[Hyades White Dwarfs]{The Hyades Cluster:  Identification of a Planetary System and Escaping White Dwarfs}

\author{B. Zuckerman$^1$, B. Klein$^1$, S. Xu(许\CJKfamily{bsmi}偲\CJKfamily{gbsn}艺)$^1$, and M. Jura$^1$}

\address{$^1$Department of Physics and Astronomy, University of California, Los Angeles, CA 90095, USA}
\eads{\mailto{ben@astro.ucla.edu}, \mailto{kleinb@astro.ucla.edu}, \mailto{sxu@astro.ucla.edu}, \mailto{jura@astro.ucla.edu}}
\begin{abstract}
Recently, some hot DA-type white dwarfs have been proposed to plausibly be escaping members of the Hyades.  We used hydrogen Balmer lines to measure the radial velocities of seven such stars and confirm that three, and perhaps two others, are/were indeed cluster members and one is not.  The other candidate Hyad is  strongly magnetic and its membership status remains uncertain.  
The photospheres of at least one quarter of field white dwarf stars are "polluted" by elements heavier than helium that have been accreted.   These stars are orbited by extended planetary systems that contain both debris belts and major planets.  We surveyed the seven classical single Hyades white dwarfs and the newly identified (escaping) Hyades white dwarfs and found calcium in the photosphere of LP 475-242 of type DBA (now DBAZ), thus  implying the presence of an orbiting planetary system.   The spectrum of white dwarf GD 31, which may be, but probably is not, an escaping member of the Hyades, displays calcium absorption lines; these originate either from the interstellar medium or, less likely, from a gaseous circumstellar disk.  If GD 31 was once a Hyades member, then it would be the first identified white dwarf Hyad with a cooling age $>$340 Myr.
\end{abstract}

Subject headings: open clusters and associations: individual (Hyades) -- stars: planetary systems -- stars: white dwarfs

\maketitle

\section{Introduction}

For decades it has been recognized that stars have evaporated from the Hyades cluster and that this process may account for the apparent paucity of white dwarf Hyads, especially white dwarfs with cooling ages $>$300 Myr (Weidemann et al 1992, Tremblay et al 2012 (hereafter TSR2012), and references therein).   
In a recent series of papers, TSR2012, Schilbach \& Roser (2012), and Roser et al (2011) use the PPMXL Catalog (Roser et al 2010) and other data to potentially substantially increase the number of identified white dwarf members of the Hyades.   Table 2 in TSR2012 presents the most recent aspects of the research; six single white dwarfs are identified as "candidate" Hyads, plus one more has a membership status of "uncertain".  All seven candidate Hyads are regarded as currently escaping from the cluster and all but one lie outside of the tidal radius of the cluster (which is $\sim$9 pc).  As described in TSR2012, the missing piece of the puzzle is a measurement of the radial velocity of the seven stars.  We measured the radial velocities by means of the hydrogen Balmer lines.  These measurements establish that three of the seven candidates are very likely to be true (escaping) Hyades members; one other and, less likely, a second, may also be escaping members. 

The Hyades contains seven "classical" single white dwarfs.  With high resolution spectroscopy, we observed these and the seven candidate white dwarf Hyads mentioned above to search for evidence of accretion of rocky material from surrounding planetary systems, should they exist.  The frequency of planetary systems in the Hyades bears on some major questions in astronomy.  For example, based on the transit technique, planetary systems in globular clusters appear to be far less common than at disk stars (Gilliland et al 2000). Is this because of the low metallicities of globular clusters or is a paucity of planetary systems generally characteristic of rich long-lasting clusters?  (The Hyades is slightly metal-rich.)   

In Section 4 we review earlier results on searches for planets and for dusty debris disks in orbit around stars in rich open clusters.  Searches via precision radial velocities and via transits are sensitive to planets within a few AU of stars.  One anticipates that these close-in regions would not be especially disturbed by encounters with other cluster members.  Rather, it is the outer planetary regions -- those sampled by direct infrared imaging, infrared photometry of cool dusty debris disks, and white dwarf accretion of rocky objects -- that typically would be most affected.  Thus, these three techniques are of special interest for evaluation of the relative nature and frequency of planetary systems in long-lived clusters and in the field. 

In Section 2 we present details of our observations and analysis and in Sections 3 and 4 consider escaping white dwarf Hyads and evidence for planetary systems in the Hyades.
Whereas our Hyades white dwarf study utilizes optical spectroscopy, Farihi et al
(2013) report an ultraviolet spectroscopic investigation of two classical white dwarf Hyads.

\section{Observations and Analysis}

We used the HIRES echelle spectrometer (Vogt et al. 1994) on the Keck I telescope at Mauna Kea Observatory in Hawaii.  Spectra were obtained during an observing run on the UT nights of 28 and 29 October 2012.  The blue cross disperser was combined with a 1.15'' slit and the full wavelength range between 3130 and 5940 \AA\ was covered with a resolution of $\sim$40,000.   Exposures of a ThAr lamp were taken throughout both nights and used for the wavelength dispersion solution.  Two radial velocity standards, HD 210667 and HD 31253 (Nidever et al. 2002), were both observed each night for absolute radial velocity calibration of the data sets.  This amounted to subtracting 1 km s$^{-1}$ from the raw velocities measured using the ThAr lamp lines.   Approximate flux calibration of the spectra  (Figures 1-3) was obtained via use of the flux standards BD+28 4211 and G191-B2B.  Spectra were reduced using both the IRAF and MAKEE software packages.  The integration time on a given white dwarf was either 40 or 60 minutes.

Figure 1 presents photospheric Ca II H- and K-lines in Hyad LP 475-242 and Figure 2 displays the same two lines in GD 31 (which may be an escaped Hyades member, see Section 3.1).   These lines in GD 31 may be of either interstellar or circumstellar origin; for reasons outlined in Section 3.3, interstellar is more likely.

With the exception of GD 31 (Section 3.1) data for the basic properties of the white dwarfs (log g, T$_{eff}$, and mass) can be found in Table 1 in TSR2012.  We use these masses and log g to calculate stellar radii and gravitational redshifts as listed in Table 1 of the present paper.   
Synthetic spectra were computed with TLUSTY \& SYNSPEC (Hubeny \& Lanz 1995).  For hydrogen dominated white dwarfs we used TLUSTY with  log g and T$_{eff}$ to calculate a pure, plane-parallel, hydrogen atmosphere in local thermodynamic equilibrium.  SYNSPEC is used to compute synthetic spectra for a given atmospheric structure. Atomic data are from the VALD (Kupka et al. 1999). The calcium upper limits are derived by comparing the equivalent width (EW) of the Ca II K-line in the spectra with the model.

For LP475-242 we use stellar parameters from Tables 1 and 2 and [H/He] =  -4.68 from Bergeron et al (2011), and derive a logarithm of the calcium to helium ratio by number [Ca/He] of -9.2.  This is typical of the ratios for polluted field DB stars given in Table 1 of Zuckerman et al (2010).  The (conservative) error estimate for [Ca/He] of 0.2 dex given in Table 2 is obtained from uncertainties in effective temperature (362 K), log g (0.09 dex), [H/He] (0.06 dex), and in measured K-line EW (8 m\AA). 

The stellar radial velocity, V, of LP 475-242 is given in the third column of Table 1.  The velocity is determined from five photospheric lines; the Ca II H- and K-lines and three strong helium lines (4471.5, 4921.9, 5875.6 \AA).  The difference between the velocity derived from the two Ca lines and, separately, from the three He lines is only 0.8 km s$^{-1}$.  The V listed in Table 1 is the average of the velocities obtained from Ca and He.  The listed uncertainty $\Delta$V is half the total spread in the velocities measured from the five lines.

All the other Table 1 stars are type DA; we used the H$\beta$ and H$\gamma$ lines to measure the radial velocities (Figure 3).  In some stars the H$\beta$ line displays a narrow non-LTE core (Figure 3) similar to those seen most often in H$\alpha$ lines in DA white dwarfs (e.g., Greenstein et al 1977; Reid 1996).  H$\delta$, H$\epsilon$, and H$\zeta$ lines are also present in our spectra, but mainly due to less precise line core profiles along with line wing asymmetries, their line centers are difficult to measure, and often result in different radial velocities from H$\beta$ and H$\gamma$. 
Two of the authors independently measured the radial velocities of the H$\beta$ and H$\gamma$ lines in the 11 non-magnetic DA white dwarfs listed in Table 1.  
Each of the two authors used IRAF's splot routine to calculate radial velocities through voigt profile fitting of the line core centers.  
In those few cases where it was apparent that the voigt function did not provide a good fit of the line core (perhaps due to lack of nearby continuum 
for the function fitting), either or both of the authors estimated by eye the location of the deepest portion of the line core and used that value for 
the line center velocity.

We could not use H$\alpha$ because it was not included in the HIRES wavelength coverage.  Measurement of H$\alpha$ line cores may improve the accuracy of the radial velocities presented in Table 1.   Our estimated uncertainties in radial velocity ($\Delta$V) are listed in the fourth column.  The $\Delta$V represent half the total spread in the four measured velocities for a given star (two persons measuring the velocities from H$\beta$ and H$\gamma$).   Therefore the listed $\Delta$V is likely to represent the maximum uncertainty in the radial velocity of a given star.

For completeness, we mention three previous papers that report radial velocities for some Table 1 stars.  Reid (1996) measured H$\alpha$ and sometimes H$\beta$ in the seven classical single white dwarf Hyads.  His H$\alpha$ velocities agree within 0.5 to 3.5 km s$^{-1}$ with our Table 1 velocities with the exception of HZ 14 where he gives an 
H$\alpha$ velocity of 68.6 km s$^{-1}$.  However, his H$\beta$ velocity agrees with ours and disagrees with his H$\alpha$ velocity.   We note that our H$\beta$ velocity for HZ 14 has a relatively large uncertainty ($\Delta$V).  If we shade our listed kinematic velocity (V$_{kin}$; Table 1 and Section 3.1) to the blue to give some weight to Reid's H$\alpha$ velocity, then V$_{kin}$ would agree even better with the TSR2012 model velocity (V$_{mod}$; Table 1 and Section 3.1) than is now listed in Table 1.

For GD 31, Zuckerman et al (2003) give a radial velocity of 85 km s$^{-1}$ for H$\beta$, in reasonable agreement with the Table 1 velocity, V, for this star, especially considering its large $\Delta$V.  If some weight is given to the Zuckerman et al (2003) velocity, then this would increase the discrepancy between V$_{kin}$ and V$_{mod}$ in Table 1.

As we note in Section 3.1, for the seven classical white dwarf Hyads, our measured kinematic radial velocities agree well and with no systematic offset from the model predicted velocities of TSR2012.   Falcon et al (2010) give radial velocities for seven Table 1 stars based on measurements of H$\beta$ and H$\alpha$ at the VLT.  These seven velocities, denoted as "apparent radial velocities", are in the Local Standard of Rest reference frame.   When converted to the heliocentric frame, within the respective errors, the Falcon et al and our Table 1 velocities are in agreement.

\section{Results}

\subsection{Candidate White Dwarf Hyads}

TSR2012 provide a clear and thorough discussion of white dwarf Hyads, those classically confirmed and also proposed candidate members.  A list of 30 initial candidate members, including 23 they reject on various grounds, appears in their Tables 1 and 2.   Actual members must have cooling times and total lifetimes not greater than the 625 Myr age of the cluster.  All but one of the proposed members lie outside of the tidal radius of the cluster, so their 3-dimensional velocity vectors must point back to near the cluster center.  These constraints eliminate all but 7 candidate white dwarfs (see Table 2 in TSR2012).   Our Table 1 provides new radial velocity data for 6 of these 7; GD 90 is strongly magnetic with uncertain physical parameters and so remains a candidate.  

In Table 1 the right hand column, V$_{mod}$, is the kinematic radial velocity expected (S. R\"oser 2012, private communication) for true Hyades members based on the traceback model of TSR2012.
Thus the relevant kinematic velocity for membership confirmation  (V$_{kin}$) must exclude the gravitational redshift  (V$_{grav}$) which is included in the HIRES-measured velocities (V) given in column 3 of Table 1.  The major uncertainties in our measured V$_{kin}$ (listed in Table 1), are due to (1) the
difficulty of obtaining accurate velocities (V) from broad hydrogen lines, and (2)
corrections for V$_{grav}$.   Errors in the parameters (log g and mass) that propagate into errors
in V$_{grav}$ are listed in Table 1 of TSR2012 and discussed in the second paragraph of their 
Section 3.  For a representative star such as GD 52, uncertainties of $\pm$0.05 dex in log g and $\pm$0.03 dex in mass imply an uncertainty of
$\pm$4 km s$^{-1}$ in redshift; this is comparable to or greater than the uncertainty in velocity measurements ($\Delta$V, 4th column in our Table 1)
of the H-beta and H-gamma lines in most target stars.  Total velocity uncertainties for a given star may be obtained by adding in quadrature $\Delta$V and uncertainties in the gravitational redshift induced by uncertainties in log g and mass. 

Comparison of V$_{kin}$ with V$_{mod}$ for the classical Hyads (stars 2,3,5,6,8,9,10 in Table 1) validates the TSR2012 model predictions and our
estimates of errors in V$_{kin}$.   The maximum offset between our measured radial velocity and the model velocity for a classical Hyad is 4.7 km s$^{-1}$ (HZ 4).  Evaluation of membership of escaping Hyads -- white dwarfs and main sequence stars -- must take into account the residual velocity dispersion in the TSR2012 model as defined and presented in their Section 5, Figure 5, and Table 2.   One anticipates larger differences -- of at least a few km s$^{-1}$ -- between measured and model radial velocities for  candidates as compared to classical members.  Then stars \# 19, 20 and 23 appear to be true (escaping) members of the Hyades cluster, while star \# 26 clearly is  
not.  

The situation for star \# 38 (GD 31) is ambiguous.   First is the problem of identifying its effective temperature.  TSR2012 give 17,470 K which comes from Gianninas et al (2011).  A similar T$_{eff}$ =  17,306 K appears in Koester et al (2009).  However, Kepler \& Nelan (1993), Gianninas et al (2005), Mullally et al (2007), and Lajoie \& Bergeron (2007) all report T$_{eff}$ to lie in the range between 12,900 and 13,500 K.  Based on model spectra (D. Koester 2013, private communication) and log g = 8.6 (Table 1), the presence of both molecular and quasi-molecular hydrogen in the spectrum of GD 31 (Section 3.3 and Xu et al 2013b) is consistent with a temperature $\sim$13,000 K but not with temperatures near 17,000 K.   In Table 2 we adopt T$_{eff}$ = 13,700 K.  This temperature is an average of temperatures based on optical and UV spectra.  For the optical, we adopt 13,744 K which is the average of the 14,118 K temperature deduced by P.-E. Tremblay (2013, private communication) based on his most recent models, and a temperature of 13,370 K derived by D. Koester (2013, private communication) from UBVJHK photometry from SIMBAD.  From the {\it Cosmic Origins Spectrograph/Hubble Space Telescope} UV data for GD 31 (ID 12169, B. Gansicke PI), D. Koester (2013, private comm.) derives a temperature of 13,640 K.  The average of the UV and optical temperatures is 13,700 K.  We set log g = 8.67 based on an average of determinations of 8.73 and 8.62 from optical data by P.-E. Tremblay (2013, private comm.) and  D. Koester (2013, private comm.), respectively.  With this log g and T$_{eff}$, the GD 31 mass and radius (Table 1) can be obtained (Fontaine et al 2001; http://www.astro.umontreal.ca/$\sim$bergeron/CoolingModels/).  

With stellar parameters from Tables 1 and 2, the white dwarf cooling time for GD 31 is $\sim$700 Myr (http://www.astro.umontreal.ca/$\sim$bergeron/CoolingModels/).  According to the initial/final mass relationship derived by Williams et al (2009), GD 31 would have been a $\sim$5 M$_{\odot}$, late B-type, star when it was on the main sequence.   Schaller et al. (1992) compute stellar evolution models for a variety of  masses and metallicities.
For a 5 M$_{\odot}$ star with Z = 0.020, the total lifetime to the end of the early AGB phase is
computed to equal 108 Myr.   Taken at face value, the total age of GD 31 would then be $\sim$800 Myr, or about 175 Myr more than the currently preferred age of the Hyades cluster.  

The difference between the calculated kinematic and model radial velocities is relatively large (Table 1) and, if some weight is given to the GD 31 velocity measured by Zuckerman et al (2003), then the difference between V$_{kin}$ and V$_{mod}$ would be increased somewhat.  Notwithstanding this difference and the moderate age discrepancy mentioned in the preceding paragraph, given reasonable uncertainties in the various parameters and models, it remains possible, albeit unlikely, that GD 31 is an escaping cluster member.  
If so, then it likely left the Hyades a few x 10$^5$ years ago.

Star \# 24 (GD 77) is magnetic.  P. Dufour (2013, private comm.) estimated the zero-field radial velocity of GD 77 from the HIRES-measured H$\beta$ and H$\gamma$ lines for various assumed magnetic fields.  A model with a 1.6 MG field gives a tolerable fit to the line profiles, but with radial velocities from the H$\beta$ and H$\gamma$ lines that differ by 24 km s$^{-1}$.  The average of these two velocities is listed as V in Table 1.  By varying the orientation and offset of a dipole field one can likely improve the profile fits and decrease the 24 km s$^{-1}$ discrepancy;  also perhaps a dipole is too simple.  Given the
current state of magnetic white dwarf modeling, improvements of these results will take some effort.  Because few magnetic white dwarfs have been observed with the high resolution of Keck/HIRES, eventually similar such measurements may be used to calibrate and improve the models..  In any event, GD 77 might be an escaping Hyad.

A recent N-body simulation of the Hyades cluster by Ernst et al (2011) is discussed at length in Section 6.1 of TSR2012.  In the Ernst et al simulation the primary mechanism for loss of white dwarfs is the velocity kick received by a white dwarf as a consequence of asymmetrical mass loss during the preceding AGB and planetary nebula phases of stellar evolution; the velocity kick is drawn from a Maxwellian distribution with an assumed 1D velocity dispersion of 5 km s$^{-1}$, corresponding to a 3D kick with mean $\sim$8 km s$^{-1}$.  To the extent that this average kick velocity is not known or well constrained and may depend on stellar mass on the main sequence, predictions drawn from the Ernst et al calculations may not be accurate.  That said, at least for white dwarfs with cooling ages $<$340 Myr, the Ernst simulations agree with the number of escaping white dwarf we identify in Table 2.  Our results (3 to 5 confirmed escaping Hyads out of 7 checked) are also consistent with the anticipated number of contaminant field white dwarfs estimated from the Monte Carlo simulations of TSR2012 (their Section 5.1 and Table 3).

\subsection{Calcium and Magnesium Abundances}

About 1/3 of field DB white dwarfs with effective temperatures in the range between 13,500  and 19,500 K reveal detectable photospheric Ca II K-lines when 
observed with the sensitivity we obtain with Keck/HIRES (Zuckerman et al 2010).  Therefore, if early-type Hyades stars possess comparable numbers of planetary systems as A- and F-type field stars, then it is not surprising to discover that LP 475-242 -- the one Hyades DB (T$_{eff}$ = 15,120 K) -- has calcium pollution. 

According to the initial-to-final mass relationship given in Williams et al (2009), when on the main sequence, LP 475-242 had a mass $\sim$3 M$_{\odot}$.  There 
are 15 previously known metal-rich DBZ white dwarfs listed in Table 1 of Jura \& Xu (2012).  Thirteen of these had main sequence masses between about 
2 and 3 M$_{\odot}$ when on the main sequence, while two were outside of this range (one larger and one smaller).  Thus, the main sequence progenitor of 
LP 475-242 was slightly more massive than the typical DBZ progenitor, but is not unusual in this regard.

Typically, because of the greater transparency of helium atmospheres, a given fractional calcium abundance enables the build-up of a much larger Ca K-line equivalent width in comparison to the EW in a hydrogen atmosphere of comparable temperature.  Thus
detection of a K-line in the hot DAs in Table 1 requires a large [Ca/H] ratio (see Table 2) compared to that measured for the polluted DB star LP 475-242.   
Some examples of highly polluted DAZ stars with T$_{eff}$ $>$14,000 K appear in Koester et al (2005), Kawka et al (2011), G\"ansicke et al (2012), and Farihi et al (2012).   Many more hot DAZs with at least some Si pollution as measured with COS are reported by Koester et al (2012).  Given the few, hot, field DA white dwarfs known to have [Ca/H] larger than the upper limits given in Table 2 for the DA Hyads, it certainly remains possible that some of these stars have atmospheres substantially polluted with heavy elements.  In particular, the statistics presented by Koester et al (2005) for hot field DAs imply that, at our level of sensitivity, the percentage of DAZs among the DA population in the Hyades could be as large as in the field DA population.   For field DA white dwarfs the percentage of DAZs, and thus likely planetary systems, is at least 25\% (Zuckerman et al 2003).  This percentage is deduced from a study of cool DAs in which a Ca II K-line can be detected more easily than in a hot DA.

The right hand column of Table 2 presents upper limits to the [Mg/H] ratio in the four hottest white dwarf Hyads.   When compared to solar ratios -- [Mg/Ca]$_{\odot}$ = 1.2 dex -- for these four stars [Mg/H] is a more constraining abundance upper limit than is the [Ca/H] ratio.  However, [Mg/Ca] can often vary by up to $\pm$0.6 dex in extrasolar rocky objects (Jura \& Xu 2013).  Thus, depending on the unknown [Mg/Ca] ratios in VR 7 and HZ 7, the calcium abundance upper limit could be the stronger.  

\subsection{The Spectrum of GD 31}

Figure 2 displays the Ca II K- and H-lines seen in the spectrum of GD 31.  These cannot be photospheric because of the large offset between their radial velocities (11.2 km s$^{-1}$ and 10.7 km s$^{-1}$ for the K- and H-line, respectively) and that of the white dwarf (89 km s$^{-1}$).  The lines must therefore be either of interstellar or circumstellar origin.    Zuckerman et al (2003) measured a radial velocity of 12 km s$^{-1}$ for the K-line which they attributed to interstellar absorption.  Given the uncertainty in the gravitational redshift for GD 31 and the relatively large uncertainty ($\Delta$V) in its photospheric velocity (V), the velocities of the calcium lines do not disagree with the expected velocity of circumstellar lines ($\sim$15 km s$^{-1}$).   Calcium K-lines are observed in the photospheres of all white dwarfs with known circumstellar disks.  If GD 31 has a circumstellar disk, then it would be the first exception to this rule.  Because of our only moderate sensitivity to photospheric calcium in GD 31 (Table 2), an exception in this case would be plausible. 
One star in which both circumstellar and photospheric Ca K-lines are detected is DAZ WD1124-293 for which T$_{eff}$$\sim$9400 K (Debes et al 2012).

Interstellar absorption lines of Ca II as well as numerous UV resonance lines are commonly detected in stars within the solar neighborhood (e.g., Lehner et al 2003; Welsh et al 2010).
According to TSR2012, GD 31 is only $\sim$30 pc from Earth and is the closest of the 14 stars listed in our Table 1.  One would anticipate interstellar lines to be more likely in more distant stars.  In Zuckerman et al (2010) we observed 25 DB white dwarfs with comparable sensitivity to those listed in Table 1.  All 25 DB stars are further from Earth than is GD 31 and (only) four of the 25 show narrow interstellar lines.   In three cases the Ca II K-line EW is similar to that in GD 31 (14 m\AA); these three stars are at distances of 50, 59 and 81 pc.  In the fourth case, PG 2234+064, the K-line EW is 28 m\AA, but this star is 125 pc from Earth.

Interstellar column densities and radial velocities of Ca II K-lines and UV lines of Mg II and Fe II have been measured in field stars within 10 degrees or so of GD 31 (Redfield \& Linsky 2008 and S. Redfield 2013, private communication).  In the UV, the local interstellar cloud is consistently detected in the general direction of GD 31, but at a velocity around 16-19 km s$^{-1}$, not 11 km s$^{-1}$.  There are two stars, EP Eri and sigma Cet, that are closer to Earth than is GD 31 and yet show UV absorption lines of Mg II and Fe II, respectively, near 11 km s$^{-1}$.  But the anticipated Ca II K-line column density at this velocity would be substantially less than the $\sim$10$^{11}$ cm$^{-2}$ implied by the K-line seen in Figure 2.

Thus if the GD 31 calcium lines are interstellar, then there is an unusually massive interstellar cloud between it and Earth.   To assess whether  the Ca II absorption detected in the spectrum of GD 31 is produced in the interstellar medium, we have measured and listed in Table 3 the strengths of three well known ultraviolet absorption lines in the archived COS data for this star (ID 12169, B. Gansicke PI).  We measured the velocity of these UV lines to be $\sim$21 km s$^{-1}$ (Table 3), but because of known uncertainties in COS velocity measurements, the velocities are not inconsistent with those of the optical calcium lines.   It is most likely that the Table 3 UV
lines arise in the interstellar medium because absorptions from the associated excited fine structure levels (Si II 1264 \AA, C II 1336 \AA, O I 1304 and 1306 \AA) are not detected as might be expected in dense circumstellar gas.  For a specific comparison between GD 31 and a star with well-measured interstellar lines, we list in Table 3
absorption line strengths for ${\alpha}$ Vir, an early B-type star 77 pc from Earth with interstellar UV absorption lines (York \& Kinahan 1979) as well
as Ca II 3933 {\AA} (Lallement et al. 1986).  By inspection, we see that relative to the ultraviolet absorption lines, Ca II is roughly a factor of two stronger in GD 31 compared
to ${\alpha}$ Vir.  However, because there are variations in the amount of calcium depletion onto grains and because the ultraviolet lines toward ${\alpha}$ Vir
are somewhat saturated, this comparison of optical and ultraviolet line strengths suggest that the Ca absorption in the spectrum of GD 31 largely arises from interstellar gas.

In addition to the unexpectedly strong Ca II K- and H-lines, GD 31 displays yet more unanticipated lines.  The archived COS data for this star (ID 12169, B. Gansicke PI) contains absorption lines of the hydrogen molecule.   H$_2$ was first identified in our COS study of G29-38 and GD 133 (Xu et al 2013a and b).   H$_2$ transitions in both the Lyman and Werner bands are seen in the spectra of both of these $\sim$12,000 K DAZ white dwarfs.  While checking the COS spectrum of GD 31 for the lines discussed in the preceding paragraph and listed in Table 3, we noticed four 
absorption dips of low to modest signal-to-noise.  Upon aligning the spectra of GD 31 and G29-38 according to their respective heliocentric velocities, it became evident that these are four of the strongest Lyman bands of H$_2$ that are seen in the spectrum of G29-38 (see Xu et al 2013b for details).  GD 31 is thus the third DA white dwarf atmosphere identified to contain H$_2$.  As pointed out in Section 3.1, the presence or absence of H$_2$ in a white dwarf UV spectrum can break degeneracies in temperatures and gravities deduced from optical spectra.

\section{Discussion}

As noted in Section 1, a question of interest is whether extended planetary systems are disrupted in massive long-lived clusters.  Dukes \& Krumholz (2012 and references cited therein) suggest that disk disruption events are no more likely in most massive clusters than in low mass clusters because observations indicate that, for most clusters, 90\% of the stars disperse into the field within 10 Myr.  
The stellar binary fraction in the Hyades  and other long-lived clusters (Pleiades, $\alpha$ Per, Praesepe) have been considered in various papers.  Patience et al (2002) note that the distribution of semimajor axes of binary stars in these four clusters peaks at 5 AU, "a significantly smaller value" than for field stars in the solar neighborhood. 

Since disruption of extended planetary systems at stars in long-lived open clusters would plausibly be more likely than disruption of close-in planets, to test relative frequencies of occurrence it is best to use techniques that probe the outer reaches of planetary systems.  As mentioned in Section 1, these techniques include direct infrared imaging of warm young planets, infrared photometry and imaging of cool dust, and optical spectroscopy of white dwarfs.    Direct infrared searches for planets on wide orbits around stars in long-lived open clusters in not now practical because only the Hyades is close to Earth, and it is sufficiently old that even massive planets will have cooled substantially.  With the Spitzer Space Telescope, Farihi et al (2008) searched at seven Hyades white dwarfs for very massive planets but without success.  Younger rich clusters, such as the Pleiades, are sufficiently far from Earth to be beyond the reach of current direct planet imaging capabilities.   To probe well into the planet mass range, searches in long-lived clusters will require much larger telescopes than currently exist.

Spitzer 70 $\mu$m photometry of Hyades stars has been reported by Su et al (2006) and Cieza et al (2008).  Cieza et al state that cool dust was detected at two of 11 A-type Hyads but at none of 67 F-, G- or K-type Hyads.  Detection of dust emission at 24 $\mu$m from F5-type star HD 28069 was reported by Mizusawa et al (2012).  These authors do not mention that HD 28069 is a Hyades member.  They did not observe at 70 $\mu$m, so the dust temperature is unknown.   In any event, given small number statistics and considerations of sensitivity (see, e.g., Cieza et al 2008), about all that may be said at this time is that the percentage of early type stars in the Hyades cluster with cool dusty debris disks may be comparable to the percentage of field stars of similar age that possess similar disks.

To date, five techniques for studying extrasolar planetary systems have proved to be fruitful; four of these are  precision radial velocities (PRV), transits, microlensing, and direct imaging of self-luminous planets.  Because of a variety of limitations, none of these four techniques have yet proved effective as a way to address the question of ''how common are planetary systems in long-lived open clusters?''  For example, although PRV revealed no planets around $\sim$100 F- through M-type dwarfs in the Hyades cluster, the results were quite insensitive because of the �noisy� behavior of young star photospheres (Paulson et al 2004).  Via the PRV technique, Sato et al (2007) report a giant planet close to the Hyades giant star eps Tau; but the unknown vsini could allow the companion to be a brown dwarf rather than a massive planet.  Other nearby open clusters (e.g., the Pleiades, Praesepe, $\alpha$ Persei) are further away than the Hyades or younger (so even �noisier� than the Hyades) or both.  Recently, using PRV, Quinn et al (2012) found two stars in Praesepe that are orbited by "hot Jupiters" with periods of a few days. 

Study of white dwarf photospheres �polluted� with elements heavier than helium is a fifth technique that can be used to probe extrasolar planetary systems, but only out beyond a few AU; planets further in are destroyed during the AGB phase of stellar evolution. A now generally accepted model for these white dwarfs is the following:

A major planet gravitationally perturbs leftover planetesimals (e.g. asteroids) in toward the white dwarf where their orbits can intersect the tidal radius of the star.  Then the asteroid is shredded into dust and gas and eventually accreted onto the star (Debes \& Sigurdsson 2002; Jura 2003).  By study of the gas and dust that orbits some of these white dwarfs and especially by high resolution spectroscopy of the accreted material, one may deduce that at least 25-30\% of all main sequence A and F-type stars (the predecessors of the white dwarfs) possess substantial planetary systems (Zuckerman et al 2010; Zuckerman et al 2003).  This percentage is not dissimilar from the percentage of main sequence and first ascent giant stars with PRV-detected close-in planets.  For example, from a study of post-main sequence giant stars, Bowler et al. (2010) report that 1/4 or more of intermediate mass, main sequence, stars have massive planetary companions within 3 AU.  

In the present paper we have searched for heavy element pollution at 10 (or possibly 11 or 12) white dwarfs that are current or escaping members of the Hyades cluster.   Because all but one of these are hot DAs, detection of the Ca II K-line, or the Mg II 4481 \AA\ line in the hottest, requires a high fractional pollution (see [Ca/H] and [Mg/H] limits in the body and notes of Table 2).  Therefore, these DA stars place only weak constraints on the fraction of early-type Hyads with extended planetary systems.  At the high temperatures of stars in Table 2, a moderate level of pollution can be detected in only DB-type white dwarfs.   This applies only to LP 475-242; at its temperature about 1/3 of field DB white dwarfs are polluted at levels that are detectable with HIRES (Zuckerman et al 2010).   The fact that calcium is detected in LP 475-242 suggests that the percentage of early-type Hyads with extended planetary systems is probably not much smaller than in the field.

\section{Conclusions}

Tremblay et al (2012 and references cited therein) consider the white dwarf members of the Hyades cluster and suggest 7 white dwarfs that may currently be escaping from the cluster.  We measured the radial velocities of the 7 single classical white dwarf members and the 7 candidate escapees.   The measured velocities suggest that three of the candidates are indeed leaving the cluster, while one was never a cluster member.  The jury is still out on the other three candidate escapees; one is probably not a member and the other two are magnetic.   None of the classical white dwarf members have cooling ages $>$340 Myr and neither do any of the likely escaping members we have identified.  However, if GD31 is an escaping member, which appears possible, albeit rather unlikely, then its cooling age is not much less than the $\sim$625 Myr age of the cluster.

The DBA (now DBAZ) white dwarf LP 475-242 displays photospheric calcium lines indicative of a surrounding planetary system.  The spectrum of the DA white dwarf GD 31 displays narrow Ca II K- and H-lines.  These might be circumstellar, but are more likely interstellar in which case an usually substantial interstellar cloud exists in the $\sim$30 pc interval between Earth and GD 31.   Also noteworthy is the presence of molecular hydrogen in the ultraviolet spectrum of this high-gravity $\sim$13,700 K white dwarf.  This is the hottest among the three white dwarf atmospheres that are known to contain H$_2$ (Xu et al 2013b).

\vskip 0.2in 

We are grateful to Siegfried R\"oser, Elena Schilbach, and Pier-Emmanuel Tremblay for numerous helpful communications, Seth Redfield and Barry Welsh for guidance on interstellar calcium lines, Detlev Koester for his estimation of the gravity and temperature of GD 31, Patrick Dufour for his calculations of line splittings in magnetic white dwarf GD 77, Adela Kawka for communication of unpublished Ca II K-line EWs, and Boris G\"ansicke for helping to set us straight on the temperature of GD 31.  This research was funded in part by NASA  and NSF grants to UCLA.  Observations of GD 31 made with the NASA/ESA Hubble Space Telescope, were obtained from the Data Archive at the Space Telescope Science Institute, which is operated by the Association of Universities for Research in Astronomy, Inc., under NASA contract NAS 5-26555. These HST observations are associated with program \#12169 (B. G\"ansicke PI).  Most data presented herein were obtained at the W.M. Keck Observatory, which is operated as a scientific partnership among the California Institute of Technology, the University of California, and the National Aeronautics and Space Administration.   The Observatory was made possible by the generous financial support of the W.M. Keck Foundation.  We thank the Keck Observatory support staff for their assistance.  We recognize and acknowledge the very significant cultural role and reverence that the summit of Mauna Kea has always had within the indigenous Hawaiian community.  We are most fortunate to have the opportunity to conduct observations from this mountain.

\section*{References}
\begin{harvard}

\item[Bergeron, P., Wesemael, F., Dufour, P. et al. 2011, ApJ 737, 28]
\item[Bowler, B., Johnson, J., Marcy, G. 2010, ApJ 709, 396]
\item[Cieza, L., Cochran, W. \&Augereau, J.-C. 2008, ApJ 679, 720]
\item[Debes, J., Kilic, M., Faedi, F. et al. 2012, ApJ 754, 59] 
\item[Debes, J. \& Sigurdsson, S. 2002, ApJ 572, 556]
\item[Dukes, D. \& Krumholz, M. 2012, ApJ 754, 56]
\item[Ernst, A., Just, A., Berczik, P. \& Olczak, C. 2011, A\&A 536, A64]
\item[Falcon, R., Winget, D., Montgomery, M. \& Williams, K. 2010, ApJ 712, 585]
\item[Farihi, J., Becklin, E. \& Zuckerman, B. 2008, ApJ 681, 1470]
\item[Farihi, J., G\"ansicke, B. \& Koester, D. 2013, MNRAS in press (arXiv:1302.6992)]
\item[Farihi, J., G\"ansicke, B. Wyatt, M. et al. 2012, MNRAS 424, 464]
\item[Fontaine, G., Brassard, P. \& Bergeron, P. 2001, PASP 113, 409]
\item[G\"ansicke, B., Koester, D. Farihi, J. et al 2012, MNRAS 424, 333]
\item[Gianninas, A., Bergeron, P. \& Fontaine, G. 2005, ApJ 631, 1100]
\item[Gianninas, A., Bergeron, P. \& Ruiz, M. 2011, ApJ 743, 138]
\item[Gilliland, R., Brown, T. \& Guhathakurta, P. 2000, ApJ 545, L47]
\item[Greenstein, J., Boksenberg, A., Carswell, R. \& Shortridge, K. 1977, ApJ 212, 186]
\item[Hubeny, I., \& Lanz, T.\ 1995, ApJ, 439, 875]
\item[Jura, M. 2003, ApJ 548, L91]
\item[Jura, M. \& Xu, S. 2012, AJ 143, 6]
\item[Jura, M. \& Xu, S. 2013, AJ 145, 30]
\item[Kawka, A., et al 2011, in Planetary Systems  Beyond the Main Sequence: AIP] 
Conference Proceedings, Volume 1331, 238
\item[Kepler, S. \& Nelan, E. 1993, AJ 105, 608]
\item[Koester, D., G\"ansicke, B., Girven, J. \& Farihi, J. 2012, arXiv:1209.6036] 
\item[Koester, D., Rollenhagen, K., Napiwotzki, R. et al. 2005, A\&A 432, 1025] 
\item[Koester, D., Voss, B., Napiwotzki, R. et al 2009, A\&A 505, 441] 
\item[Kupka, F., Piskunov, N., Ryabchikova, T., Stempels, H. \& Weiss, W. 1999, A\&AS, 138, 119]
\item[Lajoie, C.-P. \& Bergeron, P. 2007, ApJ 667, 1126]
\item[Lallement, R., Vidal-Madjar, A., \& Ferlet, R. 1986, A\&A 168, 225]
\item[Lehner, N., Jenkins, E. B., Gry, C., Moos, H. W., Chayer, P., \& Lacour, S. 2003, ApJ, 595, 858]
\item[Mizusawa, T., Rebull, L. Stauffer, J. et al. 2012, AJ 144, 135]
\item[Mullally,, F., Kilic, M., Reach, W. et al 2007, ApJS 171, 206]
\item[Nidever, D., Marcy, G., Butler, R. P. et al. 2002, ApJS, 141, 503]
\item[Patience, J., Ghez, A., Reid, I. N. \& Matthews, K.2002, AJ 123, 1570]
\item[Paulson, D., Cochran, W. \& Hatzes, A. 2004, AJ 127, 3579]
\item[Quinn, S., White, R. \& Latham, D. 2012, ApJ 756, L33]
\item[Redfield, S. \& Linsky, J. 2008, ApJ 673, 283]
\item[Reid, I.N. 1996, AJ 111, 2000]
\item[R\"oser, S., Demleitner, M. \& Schilbach, E. 2010, AJ 139, 2440]
\item[R\"oser, S., Schilbach, E., Piskunov, A., Kharchenko, N. \& Scholz, R.-D. 2011, A\&A 531, A92]
\item[Sato, B., Izumiura, H., Toyota, E. et al. 2007, ApJ 661, 527]
\item[Schaller, G., Schaerer, D., Meynet, G. \& Maeder, A. 1992, A\&AS 96, 269]
\item[Schilbach, E. \& R\"oser, S. 2012, A\&A 537, A129]
\item[Su, K., Rieke, G., Stansberry, J. et al. 2006, ApJ 653, 675]
\item[Tremblay, P.-E., Schilbach, E., R\"oser S. et al 2012, A\&A 547, 99 (TSR2012)]
\item[Vogt, S. et al. 1994, Proc. SPIE 2198, 362]
\item[Weidemann, V., Jordan, S., Iben, I. Jr. \& Casertano, S. 1992, AJ 104, 1876]
\item[Welsh, B., Lallement, R., Vergeley, J.-L., \& Raimond, S. 2010, A\&AS, 510, 54]
\item[Williams, K., Bolte, M. \& Koester, D. 2009, ApJ 693, 355]
\item[Xu, S., Jura, M. \& Koester, D. 2013a, paper 22142406 presented at the Long Beach AAS meeting]
\item[Xu, S., Jura, M., Koester, D., Klein, B. \& Zuckerman, B. 2013b, ApJL 766, L18]
\item[York, D. G., \& Kinahan, B. 1979, ApJ, 228, 127]
\item[Zuckerman, B., Koester, D., Reid, I. N. \& H�nsch, M. 2003, ApJ 596, 477]
\item[Zuckerman, B., Melis, C., Klein, B., Koester, D. \& Jura, M. 2010, ApJ 722, 725]

\end{harvard}

\clearpage

\begin{table}
\caption{Classical and Proposed White Dwarf Hyades Members}
\begin{tabular}{@{}lccccccccc}
\br
Star& Star& V& $\Delta$V& log g& mass& radius& V$_{grav}$& V$_{kin}$&  V$_{mod}$\\
& \#& (km s$^{-1}$)& (km s$^{-1}$)& (cgs)& (M$_{\odot}$)& (0.01R$_{\odot}$)& (km s$^{-1}$)& (km s$^{-1}$)& (km s$^{-1}$)\\
\mr
GD 31& 38& 89.0& 5.2& 8.67& 1.02& 0.778& 83.2& 5.8& 22.5\\
GD 52& 19& 88.6& 2.1& 8.31& 0.80& 1.043& 48.7& 39.9& 31.6\\
HZ 4& 2& 79.4& 1.3& 8.30& 0.80& 1.043& 48.7& 30.7& 35.4\\
HG7-85& 20& 86.3& 2.2& 8.25& 0.76& 1.09& 44.3& 42.0& 36.2\\
LB 227& 3& 93.1& 0.4& 8.38& 0.85& 1.00& 54.0& 39.1& 36.6\\
VR 7& 5& 75.0& 1.9& 8.13& 0.70& 1.20& 37.0& 38.0& 38.2\\
VR 16& 6& 75.4& 1.5& 8.12& 0.71& 1.20& 37.0& 38.4& 38.6\\
HZ 7& 8& 73.3& 3.6& 8.11& 0.69& 1.22& 35.9& 37.4& 39.5\\
LP475-242& 9& 85.7& 1.4& 8.25& 0.74& 1.08& 44.0& 41.7& 40.0\\
HZ 14& 10& 81.4& 4.6& 8.15& 0.73& 1.20& 38.6& 42.8& 40.2\\
GD 74& 23& 68.2& 0.4& 8.07& 0.66& 1.25& 33.5& 34.7& 37.2\\
GD 77& 24& 69.8:& $>$12& 8.30& 0.80& 1.043& 48.7& 21.1:& 34.1 \\
GD 89& 26& 73.9& 0.7& 8.47& 0.91& 0.926& 62.4& 11.5& 34.2\\
GD 90& 27& & & 8.00:& 0.6:& & & & \\
\br
\end{tabular} 
\end{table}
\noindent Notes $-$  The Star \# column gives the star numbers in the left hand columns of Tables 1 and 2 in TSR2012.  Stars \# 2, 3, 5, 6, 8, 9, and 10 are classical Hyads (see Table 2).  HG7-85 is designated HS 0400+1451 in TSR2012.  Entries in the third column are HIRES-measured heliocentric radial velocities of the target stars (see also Section 2).
Excepting GD 31 (see discussion in Section 3.1), white dwarf masses and log g are from Table 1 in TSR2012 from which we calculated the white dwarf radii listed in the 7th column.  Other symbols in the header are defined in Sections 2 and 3.1 of the present paper.   Distances to all Table 1 stars are given in Table 2 of TSR2012, except that P.-E. Tremblay's most recent model of GD 31 (see Section 3.1) indicates a distance to GD 31 of 21.7$\pm$0.8 pc.

\clearpage

\begin{table}
\caption{Hyades Membership and Atmospheric Properties of Table 1 White Dwarfs}
\begin{tabular}{@{}lcccccccc}
\br
WD& Name& Type& Hyades& T$_{eff}$& Ca EW& [Ca/H(e)]& Mg EW& [Mg/H] \\
& & & Status& (K)& (m\AA)& & (m\AA)&   \\
\mr
0231-054& GD 31& DA& non-member?& 13700& $<$5 & $<$-9.7 & &\\
0348+339& GD 52& DA& new member& 14820& $<$10 & $<$-8.9 & &\\
0352+096& HZ 4& DA& classical& 14670& $<$8 & $<$-9.1 & & \\
& HG7-85& DA& new member& 14620& $<$10 & $<$-9.0 & & \\
0406+169& LB 227& DA& classical& 15810& $<$10  & $<$-8.6 & & \\
0421+162& VR 7& DA& classical& 20010& $<$10 & $<$-7.5 & $<$10 & $<$-6.7 \\
0425+168& VR 16& DA& classical& 25130& $<$8 & $<$-7.0 & $<$5 & $<$-6.7 \\
0431+126& HZ 7& DA& classical& 21890& $<$5 & $<$-7.6 & $<$10 & $<$-6.6\\
0437+138& LP 475-242& DBAZ& classical& 15120& 40$\pm$8 & -9.2$\pm$0.2 & & \\
0438+108& HZ 14& DA& classical& 27540& $<$8 & $<$-6.7 & $<$8 & $<$-6.4 \\
0625+415& GD 74& DA& new member& 17610& $<$8 & $<$-8.2 & &\\
0637+477& GD 77& DAP& new-member? & 14650& $<$8 & $<$-9.1 & &\\
0743+442& GD 89& DA& non-member& 15220& $<$5  & $<$-9.0 & &\\
0816+376& GD 90& DAP&  ???& 11000&  & & & \\
\br
\end{tabular} 
\end{table}
\noindent Notes $-$ The Ca EW column lists the equivalent width of the photospheric Ca II K-line.  LP 475-242 displays a photospheric Ca II H-line of EW $\sim$30 m\AA, see Figure 1.  GD 31 displays narrow Ca II K and H absorption lines (see Figure 2 and Section 3.3) which are either interstellar or, less likely, circumstellar.  Na D-lines in GD 31 have EW $<$15 m\AA.  Excepting GD 31, effective temperatures (column 5) are from Table 1 in TSR2012.  The [Ca/H(e)] column gives the logarithm of the ratio by number of atoms of calcium to the most abundant constituent of a white dwarf atmosphere (which is H, except for LP 475-242).   The Mg EW column lists upper limits to the equivalent width of the Mg II 4481 \AA\ line in the four hottest white dwarfs (T$_{eff}$ $>$20,000 K).  

\clearpage

\begin{table}
\caption{Measured Ultraviolet Equivalent Widths }
\begin{tabular}{@{}lccc}
\br
Line &  GD 31 & V & ${\alpha}$ Vir \\
(\AA)   & EW (m{\AA}) & (km s$^{-1}$) & EW (m{\AA}) \\
\mr
Si II 1260  & 148$^{a}$& 23.5 & 89$^{b}$ \\
O I 1302  & 186$^{a}$ & 21.8  & 89$^{b}$ \\
C II 1335 & 170$^{a}$ & 17.2  & 112$^{b}$ \\  
Ca II 3934 & 14$^{c}$ & 11.2 & 5.2$^{d}$\\
\br
\end{tabular}
\end{table}
\noindent Notes $-$ $^{a}$Our measurements from publicly archived HST/COS data  (ID 12169, B. Gansicke PI); $^{b}$York \& Kinahan (1979); $^{c}$this paper; $^{d}$Lallementet al. (1986).  The velocities, V, are for GD 31.  The Ca II H-line velocity in GD 31 is 10.7 km s$^{-1}$.  The measured FWHM of the three UV lines in GD 31 are about 33 km s$^{-1}$, and of the Ca II K-line at 3933.6 \AA, $\sim$7 km s$^{-1}$.

\clearpage

\begin{figure}
\includegraphics[width=140mm]{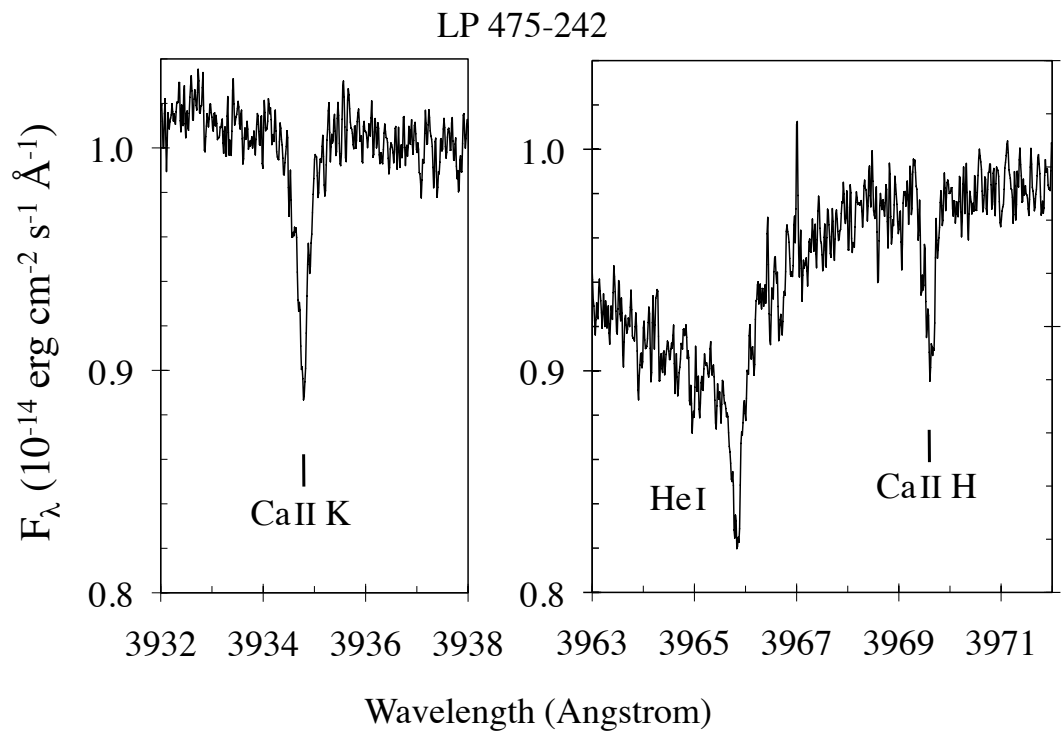}
\caption{\label{figure 1} Flux calibrated spectra of the Ca II K- and H-lines in DBAZ white dwarf LP 475-242.  The abscissa is wavelength 
in air in the heliocentric frame.  The calcium and helium lines are all photospheric.  See Section 3.2 for details.}
\end{figure}

\clearpage
\begin{figure}
\includegraphics[width=140mm]{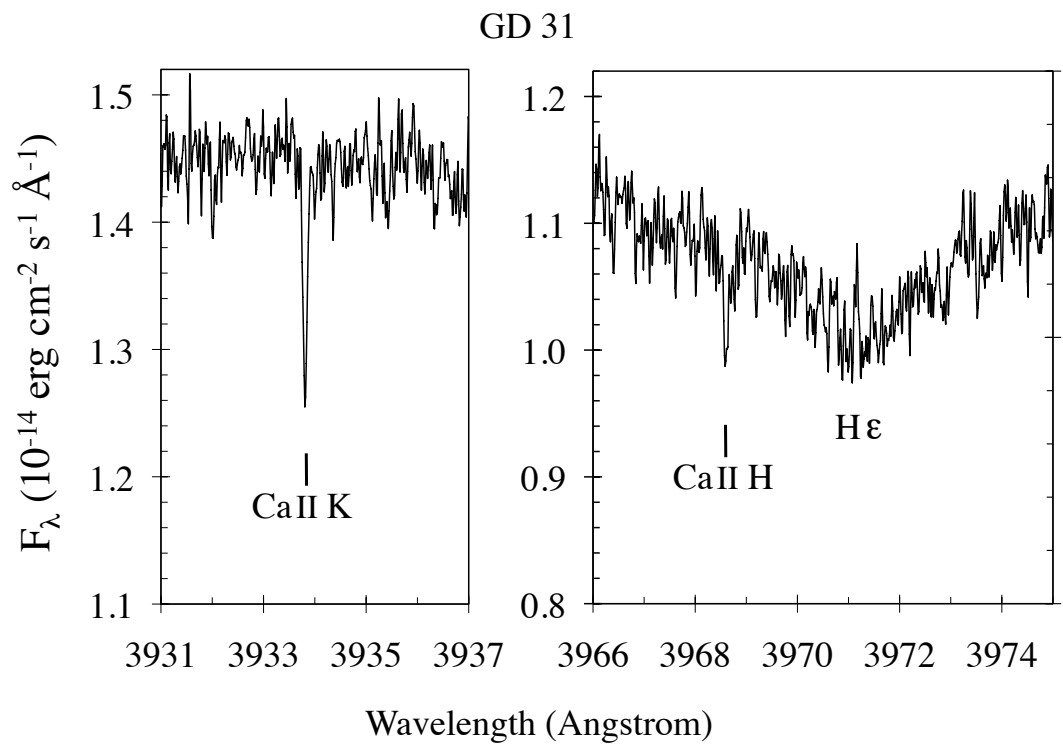}
\caption{\label{figure 2} Flux calibrated spectra of the Ca II K- and H-lines in DA white dwarf GD 31. The K-line equivalent width is 14 m\AA.   The abscissa is wavelength 
in air in the heliocentric frame. The radial velocities of the K- and H-lines are 11.2 and 10.7 km s$^{-1}$, respectively.    The calcium lines are either 
circumstellar or interstellar, probably the latter.   See Section 3.3 for details.} 
\end{figure}

\clearpage
\begin{figure}
\includegraphics[width=140mm]{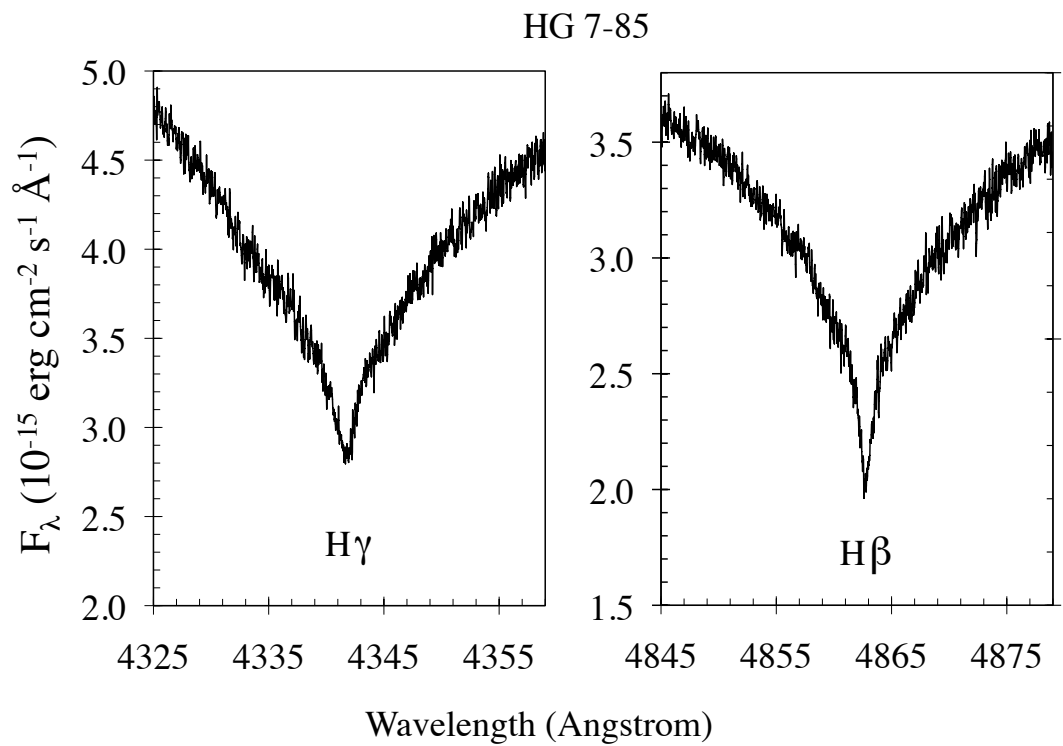}
\caption{\label{figure 3} Flux calibrated spectra of the H beta and H gamma lines in Hyades DA white dwarf HG7-85.  These lines are representative of the Balmer lines 
used for determination of the radial velocities of all stars in Table 1 with the exception of LP475-242 and GD 77. The abscissa is wavelength 
in air in the heliocentric frame.}
\end{figure}



\end{CJK}

\end{document}